\documentclass[useAMS,usenatbib]{mn2e}
\usepackage{epsfig}

\title[Can eccentric binary MSPs form by AIC of WDs?]{Can eccentric binary millisecond pulsars form by accretion induced collapse of white dwarfs?}
\author[W. C. Chen et al.]{Wen-Cong Chen$^{1,2,5}$\thanks{E-mail:
chenwc@pku.edu.cn} Xi-Wei Liu$^{3,5}$, Ren-Xin Xu$^{1}$, and
Xiang-Dong Li$^{4,5}$\\ $^1$School of Physics and State Key
Laboratory of Nuclear Physics and Technology, Peking University,
Beijing 100871, China;\\
 $^2$ Department of Physics, Shangqiu Normal University, Shangqiu 476000, China;\\
 $^3$ College of Science, Huazhong Agricultural University, Wuhan 430070, China; \\
 $^4$ Department of Astronomy, Nanjing University, Nanjing 210093, China;
 lixd@nju.edu.cn\\
 $^5$ Key Laboratory of Modern Astronomy and Astrophysics (Nanjing University), Ministry of Education, Nanjing 210093, China }
\begin{document}

\date{Accepted 2010 August 12.  Received 2010 August 11; in original form 2010 June 6}

\pagerange{\pageref{firstpage}--\pageref{lastpage}} \pubyear{2002}

\maketitle

\label{firstpage}

\begin{abstract}
Binary radio pulsars are generally believed to have been spun up
to millisecond periods (i.e. recycling) via mass accretion from
their donor stars, and they are the descendants of neutron star
low-mass X-ray binaries. However, some studies indicate that the
formation of pulsars from the accretion-induced collapse (AIC) of
accreting white dwarfs (WDs) cannot be excluded. In this work, we
use a population synthesis code to examine if the AIC channel can
produce eccentric binary millisecond pulsars (BMSPs) in the
Galaxy. Our simulated results indicate that, only when the natal
MSPs receive a relatively strong kick ($\ga100\rm km\,s^{-1}$),
can the AIC channel produce $\sim 10-180$ eccentric ($e>0.1$)
BMSPs in the Galaxy, most of which are accompanied by a Helium
star. Such a kick seems to be highly unlikely in the conventional
AIC process, hence the probability of forming eccentric BMSPs via
the AIC channel can be ruled out. Even if a high kick is allowed,
the AIC channel cannot produce eccentric BMSPs with an orbital
period of $\ga 20$ days. Therefore, we propose that the peculiar
BMSP PSR J1903+0327 cannot be formed by the AIC channel. However,
the AIC evolutionary channel may produce some fraction of isolated
millisecond pulsars, and even sub-millisecond pulsars if they
really exist.
\end{abstract}

\begin{keywords}
binaries: close -- pulsars: general -- stars: white dwarfs --
Galaxy: stellar content
\end{keywords}

\section{Introduction}
Millisecond pulsars (MSPs) are characterized by short pulse
periods ($P \la \rm 20~ ms$), low spin-down rates ($\dot{P}\sim
10^{-19}-10^{-21}\rm~ s\,s^{-1}$), old characteristic ages
($\tau=P/(2\dot{P})\sim10^{9}-10^{10}\rm ~yr$), and weak surface
magnetic fields ($B\sim10^{8}-10^{9}\rm~ G$) \citep{man04}. About
$80\%$ of the MSPs are in binaries, but only $\la 1\%$ of the
total pulsars \citep{bhat91,lori08}. Standard models proposed that
MSPs are formed in neutron star (NS) low-mass X-ray binaries
(LMXBs), in which the NS has accreted the mass and angular
momentum from its companion, and been spun up to a short spin
period \citep{alpa82,taur06}. Meanwhile, mass accretion onto the
NS induces its initial magnetic field ($B_{0}\sim
10^{12}-10^{13}~\rm G$) to decline to $B\sim 10^{8}-10^{9}~\rm G$
\footnote{As an alternative mechanism of accretion-induced field
decay, the magnetic field of the NS may be screened or buried by
the accreted material \citep{taam86,cumm01}.}. When the mass
transfer ceases, a binary millisecond pulsar (BMSP) is formed.

However, the origin of the MSPs still presents some controversial
puzzles. Firstly, can the known NS LMXBs evolve into the observed
MSPs both in the Galactic disk \citep{kulk88}? Statistical
analysises show that the birthrate of LMXBs is $\sim1-2$ order of
magnitudes lower than that of MSPs \citep{cote89,lori95}.
Secondly, it is hard to understand how isolated MSPs in the
Galactic disk were formed via the standard recycling scenario.
Although evaporation of the donor stars by the high energy
radiation of the MSPs might account for the origin of the isolated
MSPs, the observed time scales for evaporating a companion star
seem too long. Thirdly, in standard model the strong tidal effects
operating in the binary during the mass transfer serves to
circularize the orbit. However, the discovery of the eccentric
($e=0.44$) BMSP PSR J1903+0327 in the Galactic plane has
challenged the standard scenario \citep{cham08}.

As an alternative evolutionary channel, BMSP may originate from
the accretion-induced collapse (AIC) of an accreting white dwarf
(WD) \citep{mich87}. \citet{ivan04} suggested that AIC may occur
by thermal time-scale mass transfer in such binaries with orbital
periods of a few days. Once the accreting ONe WD grows to the
Chandrasekhar limit, the electron capture process may induce
gravitational collapse rather than type Ia supernova. As a result
of angular momentum conservation and magnetic flux conservation,
an MSP with rapid spin and low magnetic field may be produced. To
account for its observed characteristics, the bursting pulsar GRO
J1744-28 was suggested to be originated from the AIC of a massive
ONe WD \citep{para97,xu09}. Recently, the estimated birthrates of
MSPs by population synthesis calculations show that the
often-neglected AIC channel cannot be ignored \citep{hurl10}.

In the AIC channel the puzzles in the standard recycling model
mentioned above might disappear. Firstly, the AIC of accreting WDs
has been raised to interpret the birthrate discrepancy
\citep{bail90}, and the AIC process may be associated with the
$r$-process nucleosynthesis of heavy (baryon number $A>130$)
nuclei \citep{qian03}. Secondly, a kick velocity caused by
asymmetric collapse may be produced during AIC of WDs. An
appropriate kick can disrupt the binary system, and results in the
birth of isolated MSPs. Otherwise, the binary survives and an
eccentric BMSP is formed.

\citet{ferr07} argued that the AIC channel can form BMSPs of all
of the observed types. \citet{cham08} proposed that the AIC of a
massive and rapidly rotating WD could produce the observed orbital
parameters of PSR J1903+0327. In this work, employing the binary
population synthesis approach we attempt to investigate if the AIC
evolutionary channel can produce a population of eccentric BMSPs,
as well as isolated MSPs. In Section 2, we describe the population
synthesis approach and the evolution model of MSPs. The simulated
results by population synthesis are given in Section 3. Finally,
we present a brief discussion and summary in Section 4.

\section{Input physics}
\subsection{Population synthesis}

Using an evolutionary population synthesis  based on the rapid
binary star evolution (BSE) code \citep[e.g.][]{hurl00,hurl02}, we
attempt to study the statistical properties (such as the birth
rate, total number, the distributions of orbital period and
eccentricity) of MSPs formed via the AIC process of a massive ONe
WD in the Galaxy. In calculation, we consider the evolution of
single stars with binary-star interactions, which includes the
mass transfer and accretion via stellar winds and Roche lobe
overflow, common envelope (CE) evolution, supernovae and AIC kick,
tidal friction, and orbital angular momentum loss containing
gravitational wave radiation and magnetic braking.

All stars are assumed to born in binary systems, and start with
zero eccentricities and a solar metallicity ($Z=0.02$). We adopt
the following input parameters for the simulation. (1) A constant
star formation rate $S=7.6085~\rm yr^{-1}$, which corresponds to
that one binary with $M_{1}\geq 0.8~ M_{\odot}$ is born in the
Galaxy per year; (2) The primary mass distribution $\Phi({\rm
ln}M_{1})=M_{1}\xi(M_{1})$, in which the initial mass function
$\xi(M_{1})$ is given by \citet{krou93}; (3) The secondary mass
distribution $\Phi({\rm ln}M_{2})=M_{2}/M_{1}=q$, which
corresponds to a uniform distribution of the mass ratio $q=0-1$;
(4) A uniform distribution of ${\rm ln}a$ for the binary
separation $a$, namely $\Phi({\rm ln}a)=k=\rm constant$
(normalization results in $k=0.12328$) \citep[see][]{hurl02}. The
input parameter space for the primary mass $M_{1}$, the secondary
mass $M_{2}$, and the separation $a$ are set to be
$0.8-80M_{\odot}$, $0.1-80M_{\odot}$, and $3-10000R_{\odot}$,
respectively. For each initial parameters $\chi$ ($M_{1},M_{2}$,
and $a$), we set $n_{\chi}$  grid points in logarithmically space,
so and have
\begin{equation}
\delta \rm{ln}\chi=\frac{\rm{ln}\chi_{\rm max}-\rm{ln}\chi_{\rm
min}}{n_{\chi}-1}.
\end{equation}
During the evolution, if a binary appears as BMSP, it makes a
contribution to the birthrate as
\begin{equation}
\delta r=S\Phi({\rm ln}M_{1})\Phi({\rm ln}M_{2})\Phi({\rm
ln}a)\delta {\rm ln}M_{1}\delta {\rm ln}M_{2}\delta {\rm ln}a.
\end{equation}
If this system lives for a time $\Delta t$ as a member of BMSPs,
it makes a contribution to the number
\begin{equation}
\delta n=\delta r \times \delta t.
\end{equation}

During the CE evolution, the parameter $\alpha_{\rm CE}=E_{\rm
bind}/(E_{\rm orb,f}-E_{\rm orb,i})$ describes the efficiency that
the orbital energy is transferred to expel the envelope
\citep{hurl02}, here $E_{\rm bind}$ is the total binding energy of
the envelope, $E_{\rm orb,f}$, and $E_{\rm orb,i}$ are the final
and the initial orbital energy of the cores, respectively. There
exist some diversity in calculating the initial orbital energy,
hence \citet{hurl10} suggest that in this formulation $\alpha_{\rm
CE}=3$ should be consistent with $\alpha_{\rm CE}=1$ adopted by
\citet{pfah03}. Both the estimated velocity of individual pulsars
and the pulsar velocity distribution indicate that the nascent NSs
should have received a kick velocity, which may root in the
asymmetric collapse of massive progenitors of NSs
\citep{burr96,lai00}. The kick plays an important role in
determining the fate of binaries, whether surviving or disruptive.
Noticeably, it is widely accepted that the NS formed by the AIC
may receive a lower kick velocity (e.g. $\sigma_{\rm AIC}=20\rm
km\,s^{-1}$ in the standard model of Hurley et al. 2010). However,
some works presented the opposite point of view
\citep{harr75,lai01} (see section 4). In our standard model, a
relatively stronger kick dispersion for NSs by core collapse
supernovae is taken to be $\sigma_{\rm CC}=265\rm km\,s^{-1}$
\citep{hobb05} \footnote{To match the observed intrinsic ratio
between disrupted recycled pulsars and double NSs, the simulated
results by population synthesis show that the natal kick of NS
formed in close interacting binaries is $\sim170\rm km\,s^{-1}$
\citep{belc10}.}, while $\sigma_{\rm AIC}=150\rm km\,s^{-1}$ (a
ultra-high kick diepersion) for NSs via the AIC. To study the
influence of AIC kick on the formation of MSPs, other kick
dispersion such as $\sigma_{\rm AIC}=100,50\rm km\,s^{-1}$ are
also included. Unless we particularly mention, the BSE input
parameters in table 3 of \citet{hurl02} are used.

\subsection{Evolution of MSPs}
When the mass of the accreting WD reaches the Chandrasekhar limit,
the process of electron capture may dominate, and induce the WD to
collapse to be an NS or explode as a type Ia supernova
\citep{cana80,iser83}, depending on the WD mass and the accretion
rate. \citet{nomo91} pointed out that, if the accretion rate
$\dot{M}_{\rm wd}>10^{-8}~M_{\odot}\,\rm yr^{-1}$, an ONe WD with
an initial mass of $1.2~M_{\odot}$ is very likely to collapse to
be an NS rather than experience type Ia supernova explosion. By
2.5-dimensional radiation-hydrodynamics simulations of the AIC of
WDs, \citet{dess06} proposed that the spin period of the natal NSs
are 2.2 - 6.3 ms. Therefore, in our calculations all NSs formed by
AIC are assumed to be rapidly rotating objects with millisecond
period, and low magnetic field. In the input parameters, the
initial magnetic field strengths of the MSP are chosen from a
lognormal distribution of mean 9 and standard deviation 0.4, and
its initial spin periods $P_{\rm i}$ are distributed uniformly
between 1 and 10 ms. If the donor star fills its Roche lobe after
the AIC, the NS should appear as an X-ray pulsar
\citep{stai04,fre09}, and radio emission is suppressed. However,
even if for a detached binary with a He star or a massive
companion, accretion may occur due to capture of the stellar
winds, which may strongly influence the radio emission of MSPs.
Based on the simplified version of the theoretical model given by
\citet{davi81}, we consider the MSP's spin evolution as follows.

(i) Radio pulsar phase: After the MSP birth via AIC, it appears as
a radio pulsar due to the rapid rotation, which causes its
radiation to be strong enough to expel the winds coming from the
He star beyond the radius of the light cylinder $r_{\rm lc}$ or
the Bondi accretion radius $r_{\rm acc}$\citep{dai06}. As a result
of magnetic dipole radiation, the spin angular momentum loss rate
of the MSP is
\begin{equation}
\dot{J}_{\rm m}=-\frac{2\mu^{2}\Omega^{3}}{3c^{3}},
\end{equation}
where $\Omega$, and $\mu=10^{30}\mu_{30}\rm G\,cm^{3}$ are the
angular velocity, and the magnetic dipole moment of the MSP,
respectively.

Considering the interaction between the magnetic field of the MSP
and the wind material from the donor star, \citet{davi81} proposed
that, if the spin period $P$ of the NS is greater than either
\begin{equation}
P_{\rm ac}=1.2\dot{M}_{15}^{-1/4}\mu_{30}^{1/2}v_{8}^{-1/2}\,\rm
s,
\end{equation}
or
\begin{equation}
P_{\rm
ab}=0.8\dot{M}_{15}^{-1/6}\mu_{30}^{1/3}v_{8}^{-5/6}\left(\frac{M}{M_{\odot}}\right)^{1/3}\,\rm
s,
\end{equation}
the radio phase stops. Here $\dot{M}=10^{15}\dot{M}_{15}\rm
g\,s^{-1}$ is the rate of mass flow onto the NS, $v=10^{8}v_{8}\rm
cm\,s^{-1}$is the wind velocity relative to the NS, and $M$ is the
mass of the MSP.

Based on the criterion for the radio phase mentioned above, a NS
formed via AIC is assumed to be a MSP until its spin period
$P>10\rm ms$, or it crosses the so-called "death line", i.e.
$B_{12}/P^{2}<0.17\rm G\,s^{-2}$ \citep{bhat92}, where
$B=10^{12}B_{12}\rm G$ is the surface magnetic field of the NS.

(ii) Propeller phase: With the spin-down of the NS, the light
cylinder move outwards, and the magnetosphere radius
\begin{equation}
r_{\rm m}=1.6\times 10^{8}B^{4/7}_{12}\dot{M}^{-2/7}_{18}\,\rm cm
\end{equation}
is less than $r_{\rm lc}$, and is greater than the corotation
radius $r_{\rm c}=1.5\times10^{8}(M/M_{\odot})^{1/3}P^{2/3}\,\rm
cm$. Subjecting to the centrifugal barrier, the winds material is
assumed to be ejected at $r_{\rm m}$, and exerting a propeller
spin-down torque on the NS \citep{illa75}. The spin angular
momentum loss rate via the propeller effect is given by
\begin{equation}
\dot{J}_{\rm p}=2\dot{M}r_{\rm m}^{2}\Omega_{\rm K}(r_{\rm
m})[1-\Omega/\Omega_{\rm K}(r_{\rm m})],
\end{equation}
where $\Omega_{\rm K}(r_{\rm m})$ is the Keplerian angular
velocity at $r_{\rm m}$.

(iii) Accretion phase: we neglect the influence of the winds
accretion on the spin of the MSP owing to small spin angular
momentum of wind material.

\begin{table}
\begin{center}
\caption{Model parameters for binary population
synthesis.\label{tbl-2}}
\begin{tabular}{cccc}
\hline\hline\noalign{\smallskip}
Model & $\alpha_{\rm CE}$ &  $\sigma_{\rm CC}$&$\sigma_{\rm AIC}$\\
\hline\noalign{\smallskip}
A & 3 &  265 & 150 \\
B & 1 &  265 & 150 \\
C & 3 &  265 & 100 \\
D & 1 &  265 & 100 \\
E & 3 &  265 & 50 \\
F & 1 &  265 & 50 \\
\hline\noalign{\smallskip}
\end{tabular}
\end{center}
\end{table}

\section{Results}
Based on the theoretical model described in section 2, we
calculated the evolution of $n_{\chi}^{3}$ binaries to an age of
12 Gyr using the BSE code, as well as additions for MSP evolution
which will affect the lifetime, $\delta t$, used to calculate BMSP
numbers (see Eq.~ 3). To explore the effect of input parameters,
we have calculated the evolutionary results for six models (see
Table 1), which is determined by the CE parameter $\alpha_{\rm
CE}$, the kick dispersion $\sigma_{\rm CC}$ of core collapse, and
the kick dispersion $\sigma_{\rm AIC}$ via AIC channel.  We
collect the evolutionary stages of MSPs as follows: (1) Binary
MSPs with a main sequence, red giant, or subgiant companion
(BMSP-MG); (2) Binary MSPs with a He star companion (BMSP-He); (3)
Isolated MSPs. In Table 2, we summarize the predicted numbers and
birthrates of various types of MSPs via AIC channel in the Galaxy.

To evaluate the uncertainties in the birthrates and numbers for
various types of BMSPs, in table 2 we also present the simulated
results for same model under different random number seeds and
grid points $n_{\chi}$. We find that, the predicted numbers have
considerable uncertainties, especially the numbers of BMSP-He in
models D and F for different random number seeds, as well as the
numbers of BMSP-MG in models A, C, and E for different grid
points. However, the birthrates besides BMSP-MG show only small
variations (with uncertainties $\la 20\%$). The uncertainties in
the numbers seem to originate from various random number sequences
under different seeds or grid points. The reasons are as follows.
Firstly, the random number sequence can influence the kick
received by the natal MSP, which can affect the evolutionary fate
of binaries. Secondly, the spin evolution of MSPs is related to
the distribution of the initial period and magnetic field. Both
parameters strongly depend on the random number sequence, which
can result in the different evolutionary histories of MSPs (see
Eq.~3).

In model A, we take a strong kick of $\sigma_{\rm AIC}=150~\rm
km\,s^{-1}$ for the AIC channel, and $\alpha_{\rm CE}=3$. Our
results show that there exist $\sim 360$ BMSPs, which formed via
AIC channel. In our standard model, BMSPs with a He star companion
have a number that is $\sim10-100$ times higher than that with a
main sequence, red giant, or subgiant companion, and the birthrate
also shows the same tendency \footnote{\citet{tutu96} suggested
that binaries with non-degenerate He star may be the origin of the
most vast AIC, and outnumber binaries with H main sequence star.}.
In addition, the AIC channel can result in the formation of
$\sim5.0\times 10^{5}$ isolated MSPs, which is compatible with the
estimation by observations \citep{lori95}. As a result of large
kick, $\sim100-200 $ BMSPs with an eccentric orbit ($e\ga0.1$) is
produced.

The results of model B show that, for the same kick, a lower
$\alpha_{\rm CE}$ results in fewer BMSPs. Other models also have
the same tendency. This difference originates from the influence
of $\alpha_{\rm CE}$ on the evolution of close binaries. A high
$\alpha_{\rm CE}$ can prevent coalescence during the CE stage,
significantly increasing the formation rate of BMSPs
\citep{liu06}. In model E and F, we take a low kick for AIC
channel, and find that more BMSPs are produced, whereas almost no
eccentric BMSPs are formed. In addition, most models predict that
the number of BMSPs with a RG, or SG companion (no BMSPs with MS
donor star) is less than 10 (besides model A, C, and E when
$n_{\chi}=120$), and their eccentricities close to 0.

\begin{figure*}
 \includegraphics[width=\textwidth]{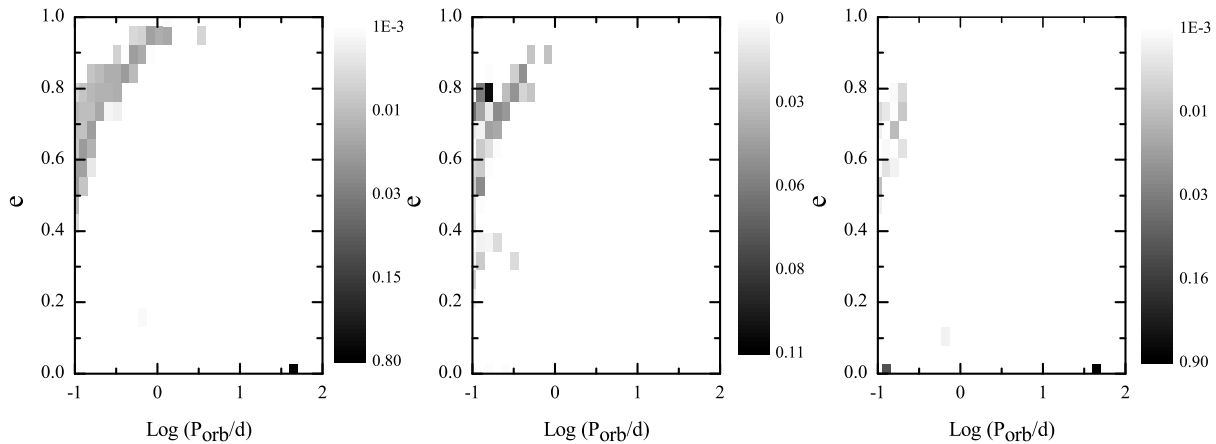}
  \caption{ Orbital period and eccentricity probability distribution of BMSPs
formed via the AIC channel for models A, B, and C, from left to
right, respectively. }
 \label{fig:fits}
\end{figure*}

In Figure 1, we plot the distribution of BMSP-He binaries in the
$e-P_{\rm orb}$ diagram for models A, B, and C under seed 1 and
$n_{\chi}=100$. It is seen that in model B most BMSP-He binaries
are in tight orbits ($P_{\rm orb}< 1\rm d $) with eccentricities
$e > 0.4$. For models A and C, the binary population seems to be
distributed into two regions. Similar to model B, the first group
have a short orbital period of $\la 1\rm d$ and a large
eccentricities of $e\ga 0.4$. While contrary to the first group,
the second are consists of BMSPs with a long orbital period of
$\ga 20\rm d$ and an eccentricities near 0.

\begin{figure*}
 \includegraphics[width=0.9\textwidth]{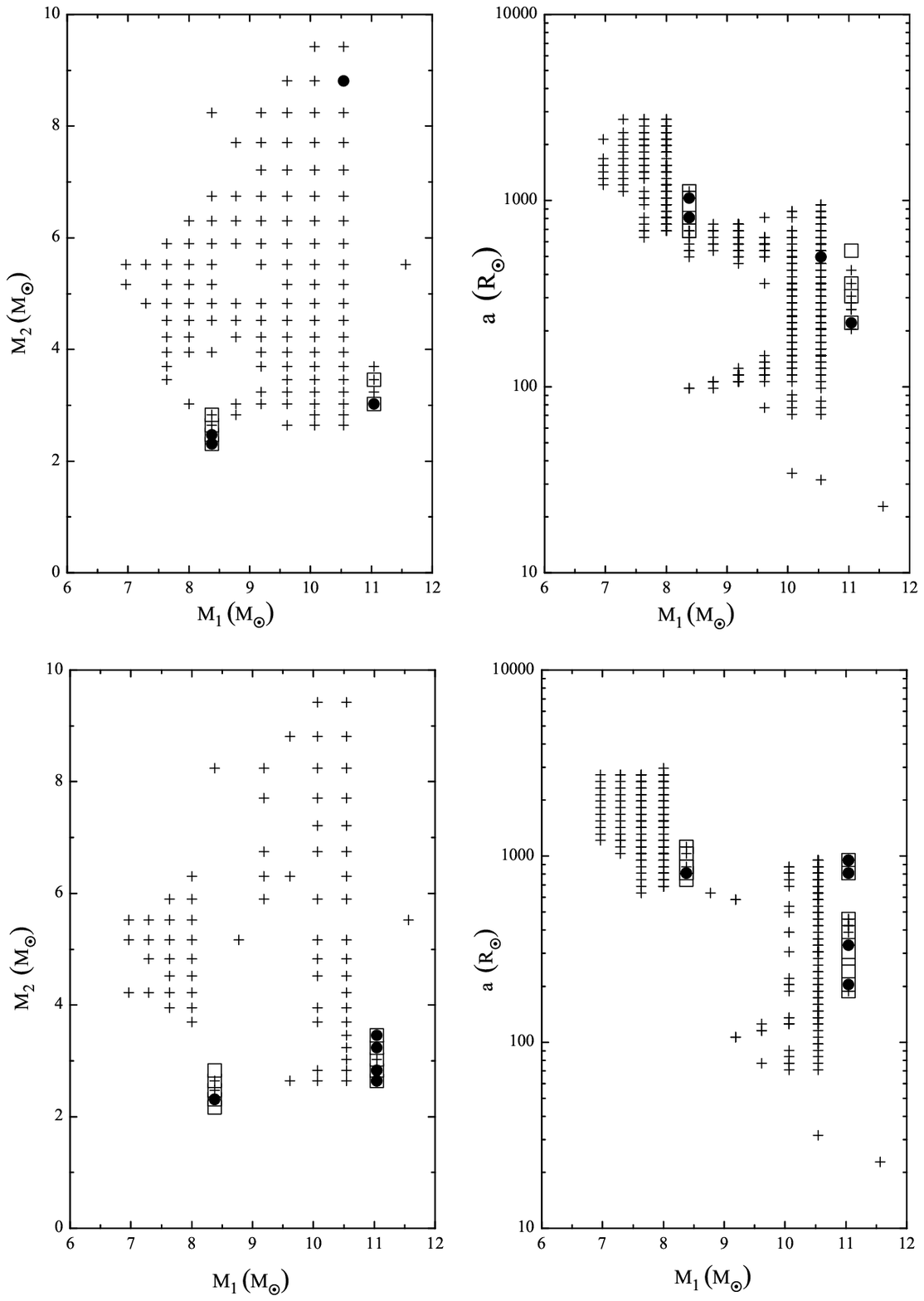}
  \caption{ Distribution of primordial binary systems forming BMSPs
in the $M_{1}-M_{2}$ and $M_{1}-a$ diagrams for models A (top
panels), and E (bottom panels). The open squares, crosses, and
solid circles correspond to the progenitors of BMSP-MG, BMSP-He,
and BMSPs with a long orbital period, respectively.}
 \label{fig:fits}
\end{figure*}

The distribution of the primordial binary systems forming BMSPs in
the $M_{1}-M_{2}$ (left) and $M_{1}-a$ (right) diagrams are shown
in Figure 2 for models A, and E. One can see that the parameter
space of the primordial binaries forming BMSP-He is wider than
that of BMSP-MG, and the latter has the rigorous initial
parameters. In addition, the initial parameter space of BMSP-He in
model A is larger than that in model E, while the tendency is
reversed for BMSP-MG. This difference results from the various
effects of the kicks on BMSP-He and BMSP-MG. The former usually
has a narrow orbit, and high kick can prevent the merger of two
components. However, for BMSP-MG with a wide orbit high kick can
lead to more binaries to be broken up during the AIC. The
difference in the initial parameter space naturally account for
the difference of birthrates. BMSPs with a long orbital period
($\ga 20\rm d$) have evolved from primordial binaries with
$M_{1}\sim 8 - 11~M_{\odot}$, $M_{2}\sim 2- 4~M_{\odot}$, and
$a\sim 100- 1000~R_{\odot}$.

\begin{table*}
\centering
\begin{minipage}{160mm}
\caption{Predicted numbers and birthrates of various types of MSPs
via the AIC channel for differential models in the Galaxy.}
\begin{tabular}{@{}lllllllllll@{}}
  \hline\hline\noalign{\smallskip}
Model&item&BMSP-He&&BMSP-MG&&isolated MSP&&BMSP&BMSP\\
&&&&&&&&($e>0.1$)&($e<0.01$)\\
  &seed,$n_{\chi}$ &number &birthrate&number & birthrate&number & birthrate&number&number\\
 \hline\noalign{\smallskip}
A&seed1,100&363& $2.3\times10^{-4}$&  4    &$4.3\times10^{-6}$&$5.0\times10^{5}$&$2.1\times10^{-4}$&107&260\\
 &seed2,100&368& $2.4\times10^{-4}$&  3    &$4.2\times10^{-6}$&$4.0\times10^{5}$&$2.1\times10^{-4}$&115&256\\
 &seed1,120&305& $2.4\times10^{-4}$&  50   &$1.1\times10^{-6}$&$4.5\times10^{5}$&$2.0\times10^{-4}$&184&170\\
\noalign{\smallskip}\hline
B&seed1,100& 21& $5.5\times10^{-5}$&  1    &$2.0\times10^{-6}$&$2.1\times10^{5}$&$9.3\times10^{-5}$&21&1  \\
 &seed2,100& 12& $5.1\times10^{-5}$&  3    &$2.0\times10^{-6}$&$2.6\times10^{5}$&$9.6\times10^{-5}$&11&3  \\
 &seed1,120& 28& $5.1\times10^{-5}$&  1    &$0.9\times10^{-6}$&$2.2\times10^{5}$&$8.1\times10^{-5}$&28&1  \\
\noalign{\smallskip}\hline
C&seed1,100&301& $2.1\times10^{-4}$&  3    &$4.6\times10^{-6}$&$4.0\times10^{5}$&$1.8\times10^{-4}$&14&290\\
 &seed2,100&503& $2.1\times10^{-4}$&  2    &$4.2\times10^{-6}$&$3.6\times10^{5}$&$1.8\times10^{-4}$&11&493\\
 &seed1,120&274& $2.2\times10^{-4}$&  62   &$1.5\times10^{-6}$&$3.9\times10^{5}$&$1.8\times10^{-4}$&58&277\\
\noalign{\smallskip}\hline
D&seed1,100& 4 & $4.6\times10^{-5}$&  2    &$2.0\times10^{-6}$&$2.0\times10^{5}$&$9.0\times10^{-5}$&4&2\\
 &seed2,100& 61& $4.7\times10^{-5}$&  2    &$4.1\times10^{-6}$&$2.4\times10^{5}$&$8.5\times10^{-5}$&8&54\\
 &seed1,120& 7 & $4.2\times10^{-5}$&  1    &$2.0\times10^{-6}$&$2.2\times10^{5}$&$7.6\times10^{-5}$&6&2\\
\noalign{\smallskip}\hline
E&seed1,100&553& $1.6\times10^{-4}$&  4    &$5.3\times10^{-6}$&$3.6\times10^{5}$&$1.7\times10^{-4}$&$<1$&556\\
 &seed2,100&610& $1.6\times10^{-4}$&  4    &$6.6\times10^{-6}$&$3.4\times10^{5}$&$1.7\times10^{-4}$&$<1$&612\\
 &seed1,120&210& $1.7\times10^{-4}$&  45   &$1.7\times10^{-6}$&$3.4\times10^{5}$&$1.7\times10^{-4}$&11  &244\\
\noalign{\smallskip}\hline
F&seed1,100&40 & $3.0\times10^{-5}$&  1    &$1.9\times10^{-6}$&$2.2\times10^{5}$&$8.3\times10^{-5}$&$<1$&40\\
 &seed2,100&320& $3.0\times10^{-5}$&  4    &$4.2\times10^{-6}$&$2.4\times10^{5}$&$8.3\times10^{-5}$&2   &322\\
 &seed1,120&52 & $3.1\times10^{-5}$&  4    &$2.4\times10^{-6}$&$2.2\times10^{5}$&$7.4\times10^{-5}$&$<1$&55\\
\noalign{\smallskip}\hline
\end{tabular}
\end{minipage}
\end{table*}

\section{Discussion and summary}
The main goal of this work is to explore if the AIC channel can
form a population of eccentric BMSPs. For a massive ONe WD that
accretes material from its donor star, when its mass reaches
Chandrasekhar limit, the process of electron capture may induce
gravitational collapse of the WD rather than a type Ia supernova
explosion. As a result of AIC, an MSP with a weak surface magnetic
field, and low spin-down rate may be formed. Same as the formation
of those NSs originated from the collapse of massive star, the
nascent NSs via AIC may also receive a kick due to the asymmetric
ejection of material or neutrinos from the newborn NSs. Our
simulated results can be summarized as follows.

1. For a high kick velocity ($150\rm km\,s^{-1}$) during the AIC,
there exist $\sim100-200$ ($\alpha_{\rm CE}=3$) or $\sim10-30$
($\alpha_{\rm CE}=1$) eccentric BMSPs in the Galaxy, accompanied
by a He star. However, almost all BMSPs are in  circular orbits
when the natal kick velocity $\la 100\rm km\,s^{-1}$.

2. The ONe WD + He star channel play an important role in forming
BMSPs. Except for model D, our predicted number and birthrate of
BMSP-He are 1-2 orders of magnitude higher than those of BMSP-MG.
In fact, this evolutionary channel has been noticed by
\citet{tutu96}. Recently, \citet{wang09a,wang09b} investigated the
evolutionary channel of CO WD + He star to type Ia supernovae, and
found that this channel can interpret type Ia supernovae with
short delay times.

3. It is difficult for the AIC channel to produce BMSPs with a MS,
RG, or SG companion, the number of which is less than 10, and
their eccentricities are near 0. If the companion of PSR
J1903+0327 is really a MS star, the probability that it originated
from the current binary can be ruled out. Both the recycled model
and the AIC channel cannot produce this peculiar BMSP with a high
eccentricity and a MS donor star, unless the natal NS had
possessed a debris disk \citep{liu09}. We expect further detailed
multi-waveband observations for PSR J1903+0327 to constrain the
properties of its optical counterpart, and hence its formation
channel.

4. All the models can produce $\sim 10^{5}$ isolated MSPs in the
Galaxy, without invoking the evaporation of donor stars by the
high energy radiation of the MSPs \citep{kluz88,rude89a,rude89b}.
In addition,  AIC of WDs may form sub-millisecond pulsars  in a
quark star regime \citep{du09}, which can hardly be formed in the
scenario of normal NSs. In particular, the AIC channel seems to be
related to the discrepancy between the birthrates of MSPs and
LMXBs, the solution of which may need additional mechanisms
forming MSPs \citep{bail90,lori95}. Certainly, it is interesting
to investigate further in this alternative model.

5. We note that the predicted numbers of various types of BMSPs
have considerable uncertainties for different random number
sequences. However, the birthrates and the numbers for BMSP-He in
a given model show little or no variation. Since BMSP-He make a
significant contribution to the total numbers of BMSPs, our
results are reliable in this respect.

Perhaps the biggest uncertainty in this work is the kick
distribution during the AIC. It strongly depend on the kick
velocity if the AIC channel can form the eccentric BMSPs. It is
usually assumed that the natal NS formed by AIC may receive a
small kick \citep{pods04,sche04,dess06,kita06}. Therefore, the
possibility forming the eccentric BMSPs via the AIC channel can be
ruled out. In addition, even if the natal NS receives a strong
kick \footnote{ The natal NS via the AIC would lose $ \sim0.2
M_{\odot}$ mass in the form of neutrinos \citep{para97}. If the
ejection of neutrinos is highly asymmetric, a large kick may be
produced. Alternatively, during AIC a large amount of mass
ejection may occur as a result of formation of a low mass bare
strange star \citep{xu05}, and a large kick velocity may be
received.}, AIC channel can also not produce the eccentric BMSPs
with an orbital period of $\ga 10$ days, and their companions are
He stars. Based on the results by binary population synthesis, we
demonstrate that the observed orbital parameters of PSR J1903+0327
cannot be produced by the AIC channel.

\section*{Acknowledgments}
We are grateful to the referee, Jarrod Hurley, for his
constructive suggestion improving this manuscript. This work was
partly supported by the National Science Foundation of China
(under grant number 10873008, 10873011, 10973002, and 10935001),
the National Basic Research Program of China (973 Program
2009CB824800), Program for Science \& Technology Innovation
Talents in Universities of Henan Province, China Postdoctoral
Science Foundation funded project, and Scientific Research
Foundation, Huazhong Agricultural University, China.

\bsp

\label{lastpage}

\end{document}